\documentclass[conference]{IEEEtran}
\usepackage[latin1]{inputenc}
\usepackage{graphicx}
\usepackage{amsmath}
\usepackage{amsfonts}
\usepackage{amssymb}
\usepackage{amsthm}
\usepackage[sans]{dsfont}
\usepackage{verbatim}
\usepackage{float}
\usepackage{threeparttable}
\usepackage{stfloats}
\usepackage{topcapt}
\usepackage{xcolor}
\restylefloat{table}

\newcommand{\av}{{\bf a}}

\newcommand{\hv}{{\bf h}}

\newcommand{\sv}{{\bf s}}

\newcommand{\wv}{{\bf w}}

\newcommand{\yv}{{\bf y}}

\newcommand{\zerov}{{\bf 0}}

\newcommand{\hbarv}{{\pmb{\hbar}}}

\newcommand{\Am}{{\bf A}}

\newcommand{\Hm}{{\bf H}}
\newcommand{\Id}{{\bf I}}

\newcommand{\Om}{{\bf O}}

\newcommand{\Qm}{{\bf Q}}
\newcommand{\Rm}{{\bf R}}

\newcommand{\Wm}{{\bf W}}

\newcommand{\Xm}{{\bf X}}
\newcommand{\Ym}{{\bf Y}}

\newcommand{\Cc}{{\cal C}}

\newcommand{\Nc}{{\cal N}}
\newcommand{\Oc}{{\cal O}}

\newcommand{\Hcb}{\pmb{\cal H}}

\long\def\symbolfootnote[#1]#2{\begingroup%
\def\thefootnote{\fnsymbol{footnote}}\footnote[#1]{#2}\endgroup} 

\begin{document}
\title{A New Low-Complexity Decodable Rate-5/4 STBC for Four Transmit Antennas with Nonvanishing Determinants}
\author{\IEEEauthorblockN{Amr Ismail, Jocelyn Fiorina, and Hikmet Sari}
\IEEEauthorblockA{Telecommunication Department, SUPELEC, F-91192 Gif-sur-Yvette,France\\
Email:$\lbrace$amr.ismail, jocelyn.fiorina, and hikmet.sari$\rbrace$@supelec.fr\\}}

\maketitle
\begin{abstract}
The use of Space-Time Block Codes (STBCs) increases significantly the optimal detection complexity at the receiver unless the low-complexity decodability property is taken into consideration in the STBC design. In this paper we propose a new low-complexity decodable rate-5/4 full-diversity 4$\times$4 STBC. We provide an analytical proof that the proposed code has the Non-Vanishing-Determinant (NVD) property, a property that can be exploited through the use of adaptive modulation which changes the transmission rate according to the wireless channel quality. We compare the proposed code to the best existing low-complexity decodable rate-5/4 full-diversity 4$\times$4 STBC in terms of performance over quasi-static Rayleigh fading channels, worst- case complexity, average complexity, and Peak-to-Average Power Ratio (PAPR). Our code is found to provide better performance, lower average decoding complexity, and lower PAPR at the expense of a slight increase in worst-case decoding complexity.
\end{abstract}


\section{Introduction}
\IEEEPARstart{S}{pace}-time coding techniques have become common-place in wireless communications standards \cite{802.11n,802.16e} as they provide an effective way to mitigate the fading phenomena inherent in wireless channels. However, the use of Space-Time Block Codes (STBCs) increases significantly the optimal decoding complexity at the receiver unless the low-complexity decodability property is taken into consideration in the STBC design. We distinguish between two decoding complexity measures, the worst-case decoding  complexity measure  and the average decoding complexity measure. The worst-case decoding complexity is defined as the minimum number of times an exhaustive search decoder has to compute the Maximum Likelihood (ML) metric to optimally estimate the transmitted symbols codeword \cite{biglieri,globcom10}, or equivalently the number of leaf nodes of the search tree if a sphere decoder is employed, whereas the average decoding complexity measure may be numerically evaluated as the average number of visited nodes by a sphere decoder in order to optimally estimate the transmitted symbols codeword \cite{burg}. It is well known that Complex Orthogonal Design (COD) codes \cite{tarokh,tirkkonen} are Single-Symbol Decodable (SSD) for general constellations. But if one considers the case of rectangular QAM constellations, the detection of each complex symbol reduces to separate detection of two real symbols which can be effectively performed via two threshold detectors (or equivalently PAM slicers) without any decoding complexity. On the other hand, the rate of square COD codes decreases exponentially with the number of transmit antennas \cite{tirkkonen}, which makes them more suitable for low-rate communications. In this paper, we address the issue of increasing the rate of the 4$\times$4 COD code at the expense of an increase in decoding complexity while preserving the coding gain and the Non-Vanishing-Determinant (NVD) property \cite{belfiore}. This property can be exploited through the use of adaptive modulation which varies the transmission rate (through the choice of the modulation order) according to the wireless channel quality. We propose a new low-complexity decodable rate-5/4 full-diversity 4$\times$4 code that encloses the rate-3/4 COD code in \cite{tirkkonen} and we provide an analytical proof that the proposed code has the NVD property.

We consider the case of square QAM constellations and compare the proposed code to the best existing low-complexity decodable rate-5/4 full-diversity 4$\times$4 STBCs in terms of performance over quasi-static Rayleigh fading channels, worst-case decoding complexity, average decoding complexity, and Peak-to-Average Power Ratio (PAPR). Our code is found to provide better performance, lower average decoding complexity and lower PAPR at the expense of a slight increase in worst-case decoding complexity that will only penalize the proposed code for high-order constellations.

The paper is organized as follows: In Section II, preliminaries on COD are provided. In Section III we introduce the new code and derive its low-complexity decodability property. In Section IV, performance comparisons by means of computer simulations are provided. We give our conclusions in Section V, and the proof of the NVD property for the proposed code is provided in the Appendix. 

\begin{table*}[b!]
\begin{threeparttable}
\setcounter{equation}{14}
\hrule
\vspace*{2.4mm}
\begin{equation}
x^{\text{ML}}_i=\textit{sign}\left(y_i'/r_{i,i}\right)\times \text{min}\Big[\big\vert 2\ \textit{round}\big(\left(y_i'/r_{i,i}-1\right)/2\big)+1\big\vert,\sqrt{M}-1 \tnote{*}\Big]\ \forall i=1,\ldots 2K. 
\label{slicer}
\end{equation}
\begin{tablenotes}
\item [*] \footnotesize{We assumed here that the underlying constellation is square such as 4-/16-QAM constellations, which is usually the case in practical scenarios}
\end{tablenotes}
\end{threeparttable}
\end{table*}

\subsection*{Notations:}
Hereafter, small letters, bold small letters and bold capital letters will designate scalars, vectors and matrices, respectively. If $\Am$ is a matrix, then $\Am^H$, $\Am^T$ and $\textit{det}[\Am]$ denote the hermitian, the transpose and the determinant of $\Am$, respectively. We define the $\textit{vec}(.)$ as the operator which, when applied to  a $m \times n$ matrix, transforms it  into a $mn\times 1$ vector by simply concatenating vertically the columns of the corresponding matrix. The $\otimes$ operator is the Kronecker product and $\delta_{kj}$ is the kronecker delta. The $\textit{sign}(.)$ operator returns 1 if its scalar input is $\geq 0$ and -1 otherwise. The $\textit{round}(.)$ operator rounds its argument to the nearest integer. The $\Re(.)$ and $\Im(.)$ operators denote the real and imaginary parts, respectively, of their argument.The $\check{(.)}$ operator when applied to a complex vector $\av$ returns $\left[\Re\left(\av^T\right),\Im\left(\av^T\right)\right]^T$ and when applied to complex matrix $\Am$ returns $\left[\Re\left( \Am^T\right),\Im\left(\Am^T\right)\right]^T$. For any two integers $a\ \text{and}\ b,\ a\equiv b \left(\text{mod}\ n\right)$ means that $a-b$ is a multiple of $n$.   

\section{Preliminaries}
\setcounter{equation}{0}
We define the MIMO channel input-output relation as: 
\begin{equation}
\underset{T\times N_r}{\Ym} =\underset{T\times N_t}{\Xm} \underset{N_t\times N_r}{\Hm} +\underset{T\times N_r}{\Wm}
\label{model}  
\end{equation} 
where $T$ is the number of channel uses, $N_r$ is the number of receive antennas, $N_t$ is the number of transmit antennas, $\Ym$ is the received signal matrix, $\Xm$ is the code matrix, $\Hm$ is the channel matrix with entries $h_{kl} \sim \Cc \Nc(0,1)$, and $\Wm$ is the noise matrix with entries $w_{ij} \sim \Cc \Nc(0,N_{0} )$. In the case of Linear Dispersion (LD) codes \cite{hassibi}, an STBC that encodes $2K$ real symbols is expressed as a linear combination of the transmitted symbols as:
\begin{equation}
\Xm=\sum^{2K}_{k=1} \pmb{\beta}_kx_k
\label{LD}
\end{equation}
with $x_k\in\mathbb{R}$ and the $\pmb{\beta}_k, k=1,...,2K$ are $T \times N_t$ complex matrices called dispersion or weight matrices that are required to be linearly independent over $\mathbb{R}$. 

The MIMO channel model can then be expressed in a useful manner by using \eqref{LD} as:
\begin{equation}
\Ym=\sum^{2K}_{k=1}\left(\pmb{\beta}_k\Hm\right)x_k+\Wm.
\end{equation} 
Applying the $\textit{vec}(.)$ operator to the above equation we obtain:
\begin{equation}
\textit{vec}(\Ym)=\sum^{2K}_{k=1}\left(\Id_{N_r}\otimes\pmb{\beta}_k\right)\textit{vec}\left(\Hm\right)x_k+\textit{vec}(\Wm).
\label{vec}
\end{equation}
where $\Id_{N_r}$ is the $N_r\times N_r$ identity matrix.\\ 
If $\yv_i$, $\hv_i$ and $\wv_i$ designate the $i$'th column of the received signal matrix $\Ym$, the channel matrix $\Hm$ and the noise matrix $\Wm$ respectively, then equation \eqref{vec} can be written in  matrix form as :
\begin{equation}
\underbrace{\begin{bmatrix}\yv_1\\\vdots\\\yv_{N_r}\end{bmatrix}}_{\yv}=\underbrace{\begin{bmatrix}\pmb{\beta}_{1}\hv_1&\dots&\pmb{\beta}_{2K}\hv_1\\\vdots&\vdots&\vdots\\\pmb{\beta}_{1}\hv_{N_r}&\dots&\pmb{\beta}_{2K}\hv_{N_r}\end{bmatrix}}_{\Hcb}\underbrace{\begin{bmatrix}x_1\\\vdots\\x_{2K}\end{bmatrix}}_{\sv}
+\underbrace{\begin{bmatrix}\wv_1\\\vdots\\\wv_{N_r}\end{bmatrix}}_{\wv}.
\end{equation}
Thus we have:
\begin{equation}
\yv=\Hcb\sv+\wv
\label{model2}
\end{equation}
A real system of equations can be obtained by separating the real and imaginary parts of the above to obtain:
\begin{equation}
\underset{2N_rT\times 1}{\check{\yv}}=\underset{2N_rT\times 2K}{\check{\Hcb}}\sv+\underset{2N_rT \times 1}{\check{\wv}}
\label{real_model}
\end{equation}
Assuming that $N_rT \geq K$, the QR decomposition of $\check{\Hcb}$ yields: 
\begin{equation}
\check{\Hcb}=\begin{bmatrix}\Qm_1 & \Qm_2\end{bmatrix} \begin{bmatrix}\Rm \\ \zerov \end{bmatrix}
\end{equation}
where $\Qm_1\in \mathbb{R}^{2N_rT\times 2K}$,$\Qm_2\in\mathbb{R}^{2N_rT \times(2N_rT-2K)}$,  $\Qm_i^T\Qm_i=\Id,\ i=1,2$, $\Rm$ is a ${2K \times 2K}$ real upper triangular matrix and $\zerov$ is a 
$(2N_rT-2K)\times 2K$ null matrix.
Accordingly, the ML estimate may be expressed as:
\begin{equation}
\sv^{\text{ML}}=\text{arg}\ \underset{\sv \in \Cc}{\text{min}}\Vert \check{\yv}-\Qm_1\Rm\sv \Vert^2
\end{equation}
where $\Cc$ is the vector space spanned by information vector $\sv$. Noting that multiplying a column vector by a unitary matrix does not alter its norm, the above reduces to:
\begin{equation}
\sv^{\text{ML}}=\text{arg}\ \underset{\sv \in \Cc}{\text{min}}\Vert \yv'-\Rm\sv \Vert^2
\label{R}
\end{equation}
where $\yv'=\Qm_1^T\check{\yv}$.
\subsection*{Complex Orthogonal Design (COD) codes:}
For COD codes, the weight matrices satisfy \cite{tirkkonen}: 
\begin{equation}
\pmb{\beta}^H_i\pmb{\beta}_j+\pmb{\beta}_j^H\pmb{\beta}_i=2\delta_{ij}\Id_{N_t}
\label{clifford}
\end{equation} 
In this case, if $\hbarv_i$ is the $i$'th column of the equivalent channel matrix $\Hcb$ in \eqref{model2}, then it is straightforward from \eqref{clifford} to prove that:
\begin{equation} 
\hbarv_k^H\hbarv_l+\hbarv_l^H\hbarv_k=0,\ \forall\ k\neq l
\end{equation}
or equivalently:
\begin{equation}
\Re\left\{\hbarv_k^H\hbarv_l\right\}=0\ \forall\ k\neq l.
\label{orthogonal}
\end{equation}
in terms of the columns of $\check{\Hcb}$ namely $\check{\hbarv}$, the above reduces to
\begin{equation} 
\check{\hbarv}^T_k\check{\hbarv}_l=0\ \forall\ k\neq l.
\label{orthogonal}
\end{equation}
The orthogonality of the columns of $\check{\Hcb}$ is inherited by the upper triangular matrix $\Rm$ in \eqref{R} which becomes simply a diagonal matrix.
\begin{table*}[t!]
\setcounter{equation}{16}
\begin{equation}
\Xm_{\textit{\textit{new}}}(\sv)=\begin{bmatrix} 
x_1+jx_2-jx_{10}e^{j\phi}&x_3+jx_4&x_5+jx_6+jx_9e^{j\phi}&-e^{j\phi}(x_7+jx_8)\\
-x_3+jx_4&x_1-jx_2-jx_{10}e^{j\phi}&e^{j\phi}(-x_7+jx_8)&-x_5-jx_6+jx_9e^{j\phi}\\
-x_5+jx_6+jx_9e^{j\phi}&e^{j\phi}(x_7+jx_8)&x_1-jx_2+jx_{10}e^{j\phi}&x_3+jx_4\\
-e^{j\phi}(-x_7+jx_8)&x_5-jx_6+jx_9e^{j\phi}&-x_3+jx_4&x_1+jx_2+jx_{10}e^{j\phi}
\end{bmatrix}
\label{newcode}
\end{equation}
\hrule
\end{table*}
In practical communication systems, the transmitted symbols are drawn from complex constellations and thus the code matrix $\Xm$ can be seen to encode $K$ complex symbols $s_i$ where $x_{2i-1}$ and $x_{2i}$ are the corresponding real and imaginary parts respectively with $i=1,\ldots,K$. The choice of the used constellation plays a key role in determining the worst-case decoding complexity, because if the complex symbols $s_i$ are drawn from a general constellation, the corresponding real and imaginary parts cannot be detected independently and thus the worst-case decoding complexity is $\Oc(M)$ where $M$ is the size of the constellation. On the other hand if the complex symbols $s_i$ are drawn from a rectangular QAM constellations, the ML decoding process of each complex symbol $s_i$ reduces to separate detection of the real and imaginary parts and the COD code can be decoded via $2K$ PAM slicers as shown in \eqref{slicer}. It is worth noting that the PAM slicer equations \eqref{slicer} require only a fixed number of simple arithmetic operations, which does not grow with the size of the rectangular QAM constellation, and therefore according to the definition of the worst-case decoding complexity they are considered of complexity $\Oc(1)$. 

\section{The Proposed Code}
From the previous section, a judicious structure for a new high-rate, low-complexity code would be to embed a COD code into a new higher rate STBC. The resulting code will enjoy the low-complexity decodability through the use of conditional detection \cite{serdar,twc}. The proposed code denoted $\Xm_{\textit{new}}$ is expressed as:
\setcounter{equation}{15}
\begin{equation}
\begin{split}
\Xm_{\textit{new}}(\sv)=&\Om(x_1,\ldots, x_6)+\\
&e^{j\phi}\left(\Rm_2 x_7+\Rm_3 x_8+\Rm_1 x_9+\Rm_5 x_{10}\right)\Rm_4 
\end{split}
\end{equation}
with $\sv=[\sv_1,\sv_2]$, $\sv_1=[x_1,\ldots,x_6]$, $\sv_2=[x_7,\ldots,x_{10}]$ and $\phi$ is chosen to maximize the coding gain. The $\Om(x_1,\ldots,x_{2(a+1)})$ matrix denotes the square COD code for the case of $2^a$ transmit antennas and $\Rm_i$ is the matrix representations of the Clifford algebra generator $\gamma_i$ with $1\leq i \leq 2a+1$ \cite{tirkkonen}. The proposed code matrix takes the form of \eqref{newcode}. It was verified through exhaustive search that taking $\phi=\frac{1}{2}\cos^{-1}(1/5)$ maximizes the coding gain and that it remains constant up to 64-QAM unnormalized constellations. An analytical proof that the coding gain of the proposed code is constant over unnormalized QAM constellations is provided in the Appendix. The corresponding upper-triangular matrix $\Rm$ takes the form below:
\setcounter{equation}{17}
\begin{equation}
\Rm=\begin{bmatrix} x&0&0&0&0&0&x&x&x&x\\
                    0&x&0&0&0&0&x&x&x&x\\
                    0&0&x&0&0&0&x&x&x&x\\
                    0&0&0&x&0&0&x&x&x&x\\
                    0&0&0&0&x&0&x&x&x&x\\
                    0&0&0&0&0&x&x&x&x&x\\
                    0&0&0&0&0&0&x&x&x&x\\
                    0&0&0&0&0&0&0&x&x&x\\
                    0&0&0&0&0&0&0&0&x&x\\
                    0&0&0&0&0&0&0&0&0&x\\\end{bmatrix}
\end{equation}
where $x$ indicates a possible non-zero position. The decoder exploits the structure of the upper triangular matrix $\Rm$ by computing the ML estimates of the orthogonal symbols $(x_1,\ldots, x_6)$ assuming that a given value of $(x_7,\ldots,x_{10})$ is transmitted. In the case of the square QAM constellations, the ML estimates of the orthogonal symbols $(x_1,\ldots, x_6)$ assuming the knowledge of $(x_7,\ldots,x_{10})$ can be obtained exactly as in \eqref{slicer} with the only difference that $y_i'$ will be replaced by $z_i=y_i'-\sum^{10}_{j=7}r_{i,j}\hat{x}_j$. Therefore, the worst-case decoding complexity of the proposed code is equal to $M^2$ as the hard PAM slicers prune the last 6 levels of the original 10 levels real valued search tree.

\section{Numerical and Simulations Results}
Table.\ref{comparison} at the top of the next page, summarizes the comparison between the proposed code and the rate-$5/4$ punctured version of the code in \cite{srinath} in terms of square QAM constellations worst-case complexity, the minimum determinant\symbolfootnote[1]{For consistency, the minimun determinant is evaluated at fixed average transmitted power per channel use for both codes}and the PAPR. The minimum determinant is defined as:
\begin{equation}
 \text{Min}\ \textit{det}=\underset{\Delta\sv \in \Delta \Cc \backslash\lbrace \zerov \rbrace}{\text{min}}\vert\textit{det}\left[\left(\Xm(\Delta\sv) \right)\right]\vert=\sqrt{\delta} 
\end{equation}
where $\Delta \sv=\sv-\sv'$, $\Delta \Cc$ is the vector space spanned by $\Delta \sv$, $\delta$ is the coding gain and the PAPR is defined as:
\begin{equation}
\text{PAPR}_n=\frac{\underset{t}{\text{max}}\vert\Xm(t,n)\vert^2}{T^{-1}\underset{t}{\sum} \mathbb{E}\lbrace\vert\Xm(t,n)\vert^2\rbrace}
\end{equation}
where $t \in \lbrace 1,\ldots T \rbrace$ and $n \in \lbrace 1,\ldots N_t \rbrace$. Due to the symmetry between transmit antennas, the subscript $n$ will be omitted. The proposed code has a higher coding gain, lower PAPR at the expense of a slight increase in worst-case detection complexity that will penalize our code for the 64-QAM constellation and above.

\begin{figure}[h!]
   \centering
      \includegraphics[scale=0.6]{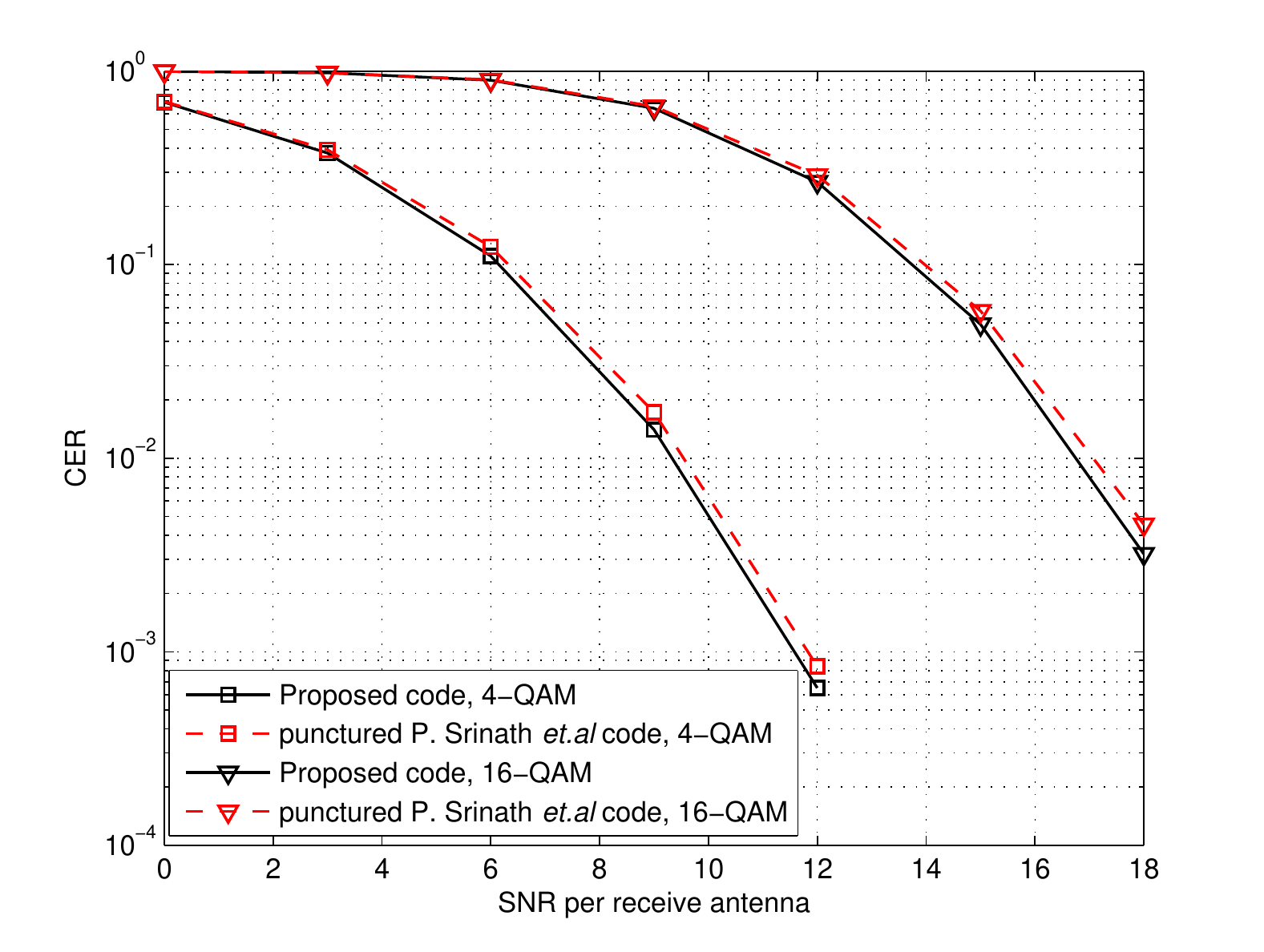} 
      \caption{CER performance for 4$\times$2 configuration and 4-/16-QAM}
      \label{CER}
\end{figure}

\begin{figure}[h!]
   \centering
      \includegraphics[scale=0.6]{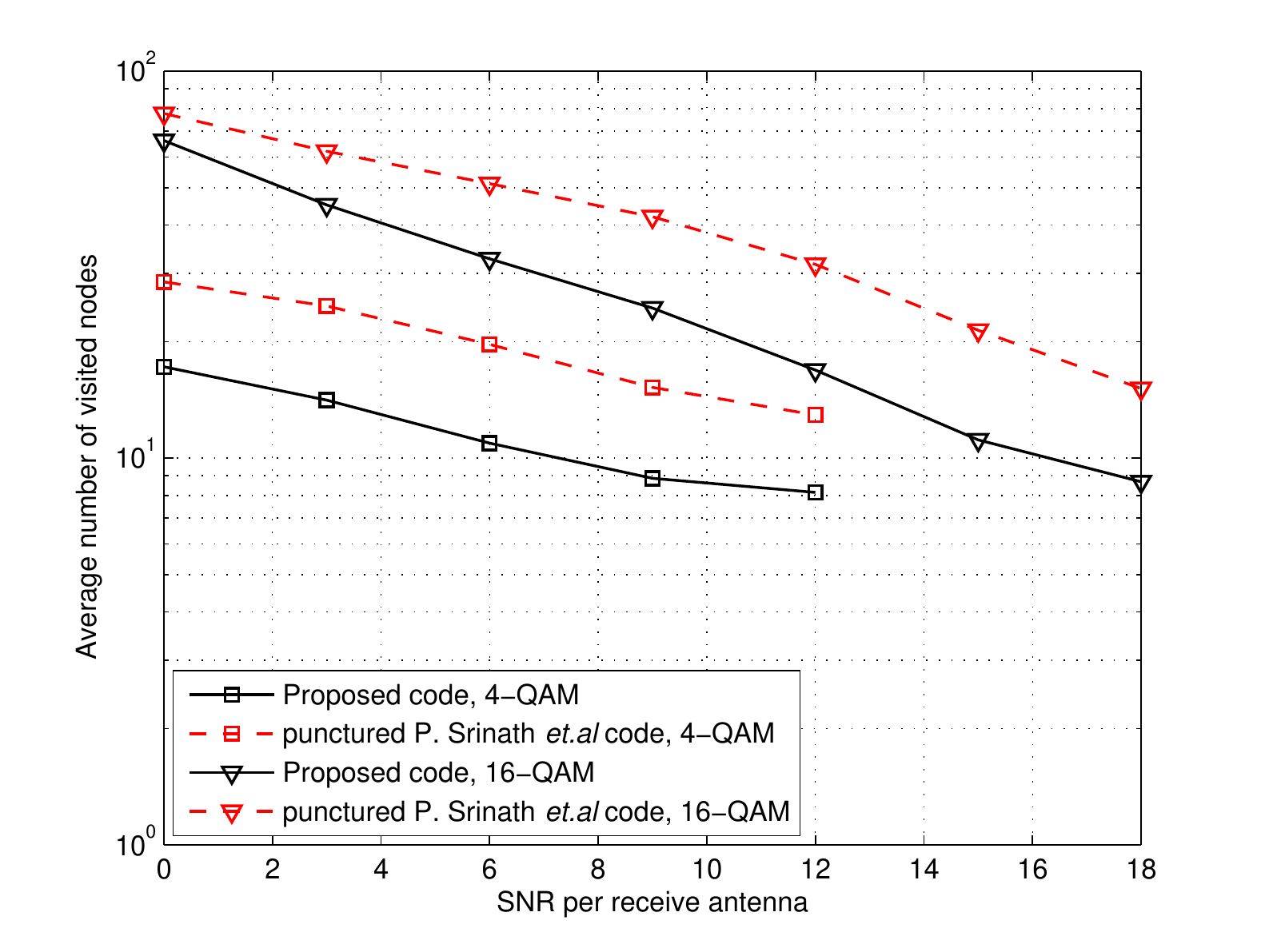} 
      \caption{Average complexity for 4$\times$2 configuration and 4-/16-QAM}
      \label{complexity}
\end{figure}

Simulations are carried out in a quasi-static Rayleigh fading channel in the presence of AWGN for $2$ receive antennas and 4-/16-QAM constellations. The ML detection is performed via a depth-first tree traversal with infinite initial radius sphere decoder. The radius is updated whenever a leaf node is reached and sibling nodes are visited according to the simplified Schnorr-Euchner enumeration \cite{SE}.  Fig.~\ref{CER} illustrates the performance comparison in terms of Codeword Error Rate (CER) of the proposed code and the rate-$5/4$ punctured version of the code in \cite{srinath}. One can notice that the proposed code offers better performance especially at high SNR. This can be interpreted by the superiority of the coding gain of our code. Furthermore, as can be seen form Fig.~\ref{complexity}, our code can be decoded with lower average complexity.

\begin{table*}
\centering
\begin{threeparttable}
\topcaption{summary of comparison in terms of worst-case complexity, Min \textit{det} and PAPR}
\begin{tabular}{|c|c|c|c|c|c|}
\hline
Code&Worst-case&Min \textit{det} ($=\sqrt{\delta}$)&\multicolumn{3}{c|}{PAPR (dB)}\\
\cline{4-6} &Complexity&for QAM constellations&QPSK&16QAM&64QAM\\
\hline
The proposed code&$M^2$&16&3.65&6.20&7.33\\
\hline
The rate-$5/4$ punctured code in \cite{srinath}&$4M^{1.5}$&12.8\tnote{*}&4.81&7.36&8.49\\
\hline
\end{tabular}
\label{comparison}
\begin{tablenotes}
\item [*] \footnotesize{This value was obtained for 4-/16-QAM as stated in \cite{srinath}}
\end{tablenotes}
\end{threeparttable}
\vspace*{2.4mm}
\hrule
\end{table*}

\section{Conclusion}
In this paper, we have introduced a new low-complexity decodable rate-5/4 full-diversity 4$\times$4 STBC that encloses the rate-3/4 COD and retains its coding gain. The coding gain is proved analytically to be constant over unnormalized QAM constellations. The proposed code is compared to the rate-5/4 punctured version of the low-complexity decodable 4$\times$4 STBC in \cite{srinath}. The proposed code showed better CER performance, lower PAPR, lower average complexity at the expense of a slight increase in worst-case detection complexity that affects our code for high-order QAM constellations. 

\section*{Appendix}
In the following, we will prove that choosing $\phi=\frac{1}{2}\cos^{-1}(1/5)$ indeed maximizes the coding gain $\delta$ (which is equal to $16^2$) for unnormalized QAM constellations and thus guarantees the NVD property for the proposed code.\\
The coding gain $\delta$ is equal to the minimum Coding Gain Distance (CGD) \cite{jafarkhani}, or mathematically:
\begin{eqnarray}
\delta &=&\underset{\sv, \sv' \in \Cc}{\underset{\sv \neq \sv'}{\text{min}}} 
\underbrace{\textit{det}\left[\left(\Xm(\sv)-\Xm(\sv')\right)^H\left(\Xm(\sv)-\Xm(\sv')\right)\right]}_{\text{CGD}(\Xm(\sv),\Xm(\sv'))}\nonumber \\
&=&\underset{\Delta\sv \in \Delta \Cc \backslash\lbrace \zerov \rbrace}{\text{min}}
\vert\textit{det}\left[\left(\Xm(\Delta\sv) \right)\right]\vert^{2}
\end{eqnarray}
where $\Delta \sv=\sv-\sv'$, $\Delta \Cc$ is the vector space spanned by $\Delta \sv$.\\
However, the code structure in \eqref{newcode} imposes:
\begin{equation*}
\underset{\Delta \sv \in \Delta \Cc \backslash\lbrace 0 \rbrace}{\text{min}}
\vert\textit{det}\left[\left(\Xm(\Delta \sv) \right)\right]\vert \leq \vert\textit{det}\left[ \Om(2,0,0,0,0,0\right]\vert=16
\end{equation*} 
As a result, the angle $\phi$ that maximizes the coding gain has to satisfy:
\begin{equation}
\vert\textit{det}\left[\left(\Xm(\Delta\sv) \right)\right]\vert \geq 16, \forall\ \Delta \sv \neq \zerov
\label{optimum}
\end{equation}
For the proposed code we have:
\begin{equation}
\begin{split}
\vert\textit{det}\left[\left(\Xm(\Delta\sv) \right)\right]\vert=&\Big\vert\left(\sum^{6}_{i=1}\Delta x_i^2 \right)^2+e^{j2\phi}b+e^{j4\phi}\left(\sum^{10}_{i=7}\Delta x_i^2\right)^2\Big\vert
\end{split}
\label{det}
\end{equation}
where $\Delta x_i=2n_i,\ n_i\in \mathbb{Z}\ i=1,\ldots,10$, and
\begin{equation*}
\begin{split}
b=2&\left(\sum^{6}_{i=1}\Delta x_i^2 \right)\left(\sum^{10}_{j=7}\Delta x_j^2\right)\\
-4\Big[&(\Delta x_7\Delta x_4-\Delta x_8\Delta x_3+\Delta x_9\Delta x_6-\Delta x_{10}\Delta x_2)^2\\
+&(\Delta x_7\Delta x_6+\Delta x_8\Delta x_2-\Delta x_9\Delta x_4-\Delta x_{10}\Delta x_3)^2\\
+&(\Delta x_7\Delta x_2-\Delta x_8\Delta x_6-\Delta x_9\Delta x_3+\Delta x_{10}\Delta x_4)^2\Big].
\end{split}
\end{equation*}
A simplification of the expression of $b$ is possible by noting that:
\begin{equation*}
\begin{split}
&\left(\sum^{6}_{i=1}\Delta x_i^2 \right)\left(\sum^{10}_{j=7}\Delta x_j^2\right)=\\
&\left(\Delta x_2^2+\Delta x_3^2+\Delta x_4^2+\Delta x_6^2\right)\left(\Delta x_7^2+\Delta x_8^2+\Delta x_9^2+\Delta x_{10}^2\right)+\\
&\left(\Delta x_1^2+\Delta x_5^2\right) \left(\Delta x_7^2+\Delta x_8^2\right) +\left(\Delta x_1^2+\Delta x_5^2\right) \left(\Delta x_9^2+\Delta x_{10}^2\right)
\end{split}
\end{equation*}
Applying the Euler's four square on the first term and the Fibonacci's two square identities on the rest of terms we obtain:
\begin{equation*}
\begin{split}
&\left(\sum^{6}_{i=1}\Delta x_i^2 \right)\left(\sum^{10}_{j=7}\Delta x_j^2\right)\\
&=(\Delta x_7\Delta x_4-\Delta x_8\Delta x_3+\Delta x_9\Delta x_6-\Delta x_{10}\Delta x_2)^2\\
&+(\Delta x_7\Delta x_6+\Delta x_8\Delta x_2-\Delta x_9\Delta x_4-\Delta x_{10}\Delta x_3)^2\\
&+(\Delta x_7\Delta x_2-\Delta x_8\Delta x_6-\Delta x_9\Delta x_3+\Delta x_{10}\Delta x_4)^2\\
&+(\Delta x_7 \Delta x_3+\Delta x_8\Delta x_4+\Delta x_9\Delta x_2+\Delta x_{10}\Delta x_6)^2\\
&+(\Delta x_7\Delta x_1+\Delta x_8\Delta x_5)^2+(\Delta x_8\Delta x_1-\Delta x_7\Delta x_5)^2\\
&+(\Delta x_9\Delta x_1+\Delta x_{10}\Delta x_5)^2+(\Delta x_{10}\Delta x_1-\Delta x_9\Delta x_5)^2\\
&=\sum^{8}_{i=1}a_i^2,\ a_i\in \mathbb{Z}. 
\end{split}
\end{equation*}
Therefore, one may write $b$ in more compact form as below:
\begin{equation}
b=2\left(\sum^{8}_{i=1}a_i^2-2(a_1^2+a_2^2+a_3^2)\right) 
\label{b}
\end{equation}
Setting  $x=e^{j2\phi}$, the discriminant $\Delta$ of the second degree equation \eqref{det} is expressed as:
\begin{equation}
\Delta=4\left(\sum^{8}_{i=1}a_i^2-2\left(a_1^2+a_2^2+a_3^2\right)\right)^2-4\left(\sum^{8}_{i=1} a_i^2\right)^2\leq 0  
\end{equation}
Consequently, the roots of equation \eqref{det} are:
\begin{equation}
\begin{split}
\lambda_{1,2}=&\frac{-b\pm j\sqrt{4\left(\sum^{6}_{i=1}\Delta x_i^2\right)^2\left(\sum^{10}_{j=7}\Delta x_j^2\right)^2 -b^2}}{2 \left(\sum^{10}_{j=7}\Delta x_j^2\right)^2},\\
&\lambda_2=\lambda_1^*,\ \vert \lambda_1\vert= \vert \lambda_2 \vert =\frac{\sum^{6}_{i=1}\Delta x_i^2}{\sum^{10}_{j=7}\Delta x_j^2}.
\end{split}
\end{equation}
For the sake of simplicity, we will denote hereafter $\sum^{6}_{i=1}\Delta x_i^2=\sigma_1$ and $\sum^{10}_{j=7} \Delta x_j^2=\sigma_2$. In the case of $\sigma_1 \neq \sigma_2$, equation.~\ref{det} can be lower bounded as below:
\begin{equation}
\begin{split}
\vert\textit{det}\left[\left(\Xm(\Delta\sv)\right)\right]\vert\Big\vert_{\underset{\sigma_1\neq\sigma_2}{\Delta \sv \neq \zerov}}&=\sigma^2_2 \Big\vert \left(x-\lambda_1\right)\left(x-\lambda_2\right) \Big\vert\\ 
&\geq \sigma_2^2\Big\vert\left(\vert x\vert-\vert\lambda_1\vert\right)\left(\vert x\vert-\vert\lambda_2\vert\right)\Big\vert\\
&=\sigma_2^2\left(1-\frac{\sigma_1}{\sigma_2}\right)^2\\
&=\left(\sigma_2-\sigma_1\right)^2\geq 16
\end{split}
\label{ineqcase1}
\end{equation}
where the latter inequality follows by substituting $\Delta x_i=2n_i,\ n_i \in \mathbb{Z}$ as we are dealing with unnormalized QAM constellations. If $\sigma_1=\sigma_2=\sigma$, equation \eqref{det} can be written as
\begin{equation}
\vert\textit{det}\left[\left(\Xm(\Delta\sv) \right)\right]\vert\Big\vert_{\underset{\sigma_1=\sigma_2}{\Delta \sv \neq \zerov}}=\left\vert 2\sigma^4\cos(2\phi)+b\right\vert
\label{case2}
\end{equation}
where $b=2\left(\sigma^2-2(a_1^2+a_2^2+a_3^2)\right)$. Taking $\cos(2\phi)=1/5$, we have to prove that:
\begin{equation}
\Big\vert \frac{2\sigma^2}{5}+b\Big\vert \geq 16 \ \forall\ \sigma_1=\sigma_2=\sigma\neq 0.
\label{ineq1}
\end{equation}
Multiplying both sides by 5 and using \eqref{b}, the above inequality becomes:
\begin{equation}
\left\vert 12 \sigma^2-20(a_1^2+a_2^2+a_3^2)\right\vert_{\left(a_1,\ldots,a_8\right)\neq \zerov} \geq  5\times 16
\end{equation}
Denoting $\tilde{a}_i=\frac{a_i}{4}$ and $\tilde{\sigma}=\frac{\sigma}{4}$ we may have:
\begin{equation}
\left\vert 12 \tilde{\sigma}^2-20(\tilde{a}_1^2+\tilde{a}_2^2+\tilde{a}_3^2)\right\vert_{\left(\tilde{a}_1,\ldots,\tilde{a}_8\right)\neq \zerov} \geq  5.
\end{equation}
The above inequality is satisfied iff:
\begin{equation}
\left\vert 3 \tilde{\sigma}^2-5(\tilde{a}_1^2+\tilde{a}_2^2+\tilde{a}_3^2)\right\vert_{\left(\tilde{a}_1,\ldots,\tilde{a}_8\right)\neq \zerov} \geq  2.
\end{equation}
However, the L.H.S of the above inequality can be considered as a special case of:
\begin{equation}
3X^2_1-5(X^2_2+X^2_3+X^2_4)\Big\vert_{\left(X_1,X_2,X_3,X_4\right)\neq \zerov } 
\label{diophantine}
\end{equation}
where $X_i \in \mathbb{Z}$. This type of equations have been extensively studied in the mathematical literature dealing with the solvability of quadratic Diophantine equations (see \cite{mordell}). Applying theorem 6  in \cite{mordell} we have:
\begin{equation}
3X^2_1-5(X^2_2+X^2_3+X^2_4) \neq 0\ \forall\ \left(X_1,X_2,X_3,X_4\right)\neq \zerov 
\end{equation} 
as $-3\times 5 \times 5\times 5\equiv 1\ (\text{mod}\ 8)$ with $3-5-5-5\equiv 4\ (\text{mod}\ 8)$.\\
Moreover, 
\begin{equation}
3X^2_1-5(X^2_2+X^2_3+X^2_4)\neq\pm 1.
\end{equation}
Otherwise, we must have:
\begin{equation}
 3X^2_1\equiv \pm 1\ (\text{mod}\ 5)
\end{equation}
which cannot be true, since the quadratic residues modulo 5 are 0,1 and 4 \cite{quadres}, thus $3X^2_1 \equiv 0,\pm 3\  \text{or}\pm 2\ (\text{mod}\ 5)$. Therefore, we can write:
\begin{equation}
3X^2_1-5(X^2_2+X^2_3+X^2_4)\Big\vert_{\left(X_1,X_2,X_3,X_4\right)\neq \zerov} \geq 2 
\end{equation}
which in turns implies:
\begin{equation}
\vert\textit{det}\left[\left(\Xm(\Delta\sv) \right)\right]\vert\Big\vert_{\underset{\sigma_1=\sigma_2}{\Delta \sv \neq \zerov}}>16
\label{ineqcase2}
\end{equation}
thus completing the proof.

\end{document}